\documentclass[10pt]{article}
\usepackage{amsmath}
\usepackage{amssymb}
\usepackage{graphicx}
\usepackage[ruled]{algorithm2e}

\def\>{\ensuremath{\rangle}}
\def\<{\ensuremath{\langle}}

\newtheorem{thm}{Theorem}[section]

\newtheorem{lem}{Lemma}[section]
\newtheorem{defn}{Definition}[section]
\newtheorem{prob}{Problem}[section]

\newtheorem{exam}{Example}[section]

\begin{document}
\newcommand{\bra}[1]{\langle #1|}
\newcommand{\ket}[1]{|#1\rangle}
\newcommand{\braket}[3]{\langle #1|#2|#3\rangle}
\newcommand{\ip}[2]{\langle #1|#2\rangle}
\newcommand{\op}[2]{|#1\rangle \langle #2|}

\newcommand{\tr}{{\rm tr}}
\newcommand{\id}{{\rm Id}}
\newcommand {\spa } {{\rm span}}
\newcommand {\supp } {{\rm supp}}
\newcommand {\A } {\mathcal{A}} 
\newcommand {\C } {{\mathbb{C}}} 
\newcommand {\F } {{\textbf{F}}}
\newcommand {\G } {{\textbf{G}}}
\newcommand {\U } {{\textbf{U}}}
\newcommand {\Uo } {{\mathcal{U}}}
\newcommand {\I } {{\textbf{I}}}
\newcommand{\hs}{\mathcal{H}}
\newcommand {\Lan} {{\mathcal{L}}}
\newcommand {\M} {{\mathcal{M}}}
\newcommand {\N } {{\mathbb{N}}} 
\newcommand {\T} {{\mathcal{T}}}
\newcommand {\me} {\mathrm{e}}
\newcommand {\mi} {\mathrm{i}}

\title{(Un)decidable Problems about Reachability of Quantum Systems}

\author{Yangjia Li and Mingsheng Ying\\
TNLIST, Dept. of CS, Tsinghua University, China\\
QCIS, FEIT, University of Technology, Sydney, Australia\\liyangjia@gmail.com}
\date{}
\maketitle
\begin{abstract}
We study the reachability problem of a quantum system modelled by a quantum automaton. The reachable sets are chosen to be boolean combinations of (closed) subspaces of the state space of the quantum system. Four different reachability properties are considered: eventually reachable, globally reachable, ultimately forever reachable, and infinitely often reachable. The main result of this paper is that all of the four reachability properties are undecidable in general; however, the last three become decidable if the reachable sets are boolean combinations without negation.

\end{abstract}

\section{Introduction}

Recently, verification of quantum systems has simultaneously emerged as an important problem from several very different fields. First, it was identified by leading physicists as one of the key steps in the simulation of many-body quantum systems \cite{CZ12}. Secondly, verification techniques for quantum protocols \cite{GPN10, AGN13} become indispensable as quantum cryptography is being commercialised. Thirdly, verification of quantum programs \cite{Yin11, YYFD13} will certainly attract more and more attention, in particular after the announcement of several scalable quantum programming languages like Quipper \cite{GV13}.

Reachability is a fundamental issue in the verification and model-checking of both classical and probabilistic systems because a large class of verification problems can be reduced to reachability analysis \cite{BK08}. Reachability of quantum systems also started to receive attention in recent years. For example, Eisert, M\"{u}ller and Gogolin's notion of quantum measurement occurrence in physics~\cite{EMG12} is essentially the reachability of null state; a certain reachability problem~\cite{SSL02} lies at the heart of quantum control theory since the controllability of a quantum mechanical system requires that all states are reachable by choosing the Hamiltonian of the system~\cite{AT12}. Reachability of quantum systems modelled by quantum automata, or more generally by quantum Markov chains, was studied by the authors \cite{YFYY13} with an application in termination analysis of quantum programs \cite{LYY12, YY12}.

This paper is a continuation of our previous work \cite{YFYY13, LYY12, YY12}, where only reachability to a single (closed) subspace of the state Hilbert space of a quantum system was considered.
In this paper, we consider a class of much more general reachability properties; that is, we use subspaces of the state space as the basic properties (atomic propositions) about the quantum system, and then reachability properties can be defined as certain temporal logical formulas over general properties, which are formalized as boolean combinations of the subspaces. The reason for using boolean combinations rather than orthomodular lattice-theoretic combinations in the Birkhoff-von Neumann quantum logic \cite{BN36} is that in applications these reachability properties will be used as a high-level specification language where boolean connectives are suitable; for example, when a physicist says that a particle will eventually enter region $A$ or region $B$, the word \textquotedblleft or\textquotedblright\ here is usually meant to be the boolean \textquotedblleft or\textquotedblright\ but not the orthomodular \textquotedblleft or\textquotedblright\ (see Example \ref{exa:QW}). The reachability properties that we are concerned with are: \begin{itemize}\item eventually reachability denoted by the temporal logic formula $\F f$; \item globally reachability denoted by $\G f$; \item ultimately forever reachability denoted by $\U f$; \item infinitely often reachability denoted by $\I f$,\end{itemize} where $f$ is a boolean combination of the subspaces of the state Hilbert space.

We use quantum automata~\cite{KW97} as a formal model for quantum systems. Then the reachability problem can be described as: decide whether or not all the execution paths of a quantum automaton satisfy $\F f$, $\G f$, $\U f$, or $\I f$. There are two reasons for adopting this model. First, it contains unitary operations so that a lot of closed physical systems can be modelled, e.g., quantum circuits. Second, without probabilistic choices (which occur in other operations such as quantum measurements and super-operators) it can be seen more clear that the reachability problem for quantum systems is essentially more difficult than that for classical systems. In fact, we note that reachability analysis is challenging in the quantum scenario, since the state space is a continuum where some techniques that have been successfully used in the classical case will become ineffective.

\subsection{Contributions of the paper}
\begin{itemize}\item We prove undecidability of the above reachability problem, even with $f$ in a very simple form containing the boolean negation. Undecidability of $\G f$ (globally reachable), $\U f$ (ultimately forever reachable) and $\I f$  (infinitely often reachable) comes from a straightforward reduction from the emptiness problem for quantum automata~\cite{BJKP05}. However, undecidability of $\F f$ (eventually reachable) requires a careful reduction from the halting problem for 2-counter Minsky machines \cite{Min76}. In particular, a novel strategy is introduced in this reduction to simulate a (possibly irreversible) classical computation using a quantum automaton which is definitely reversible. These undecidability results present an impressive difference between quantum systems and classical systems because the reachability properties considered in this paper are decidable for classical systems.

\item We prove decidability of the reachability problem for $\G f$, $\U f$, and $\I f$ with $f$ being positive; that is, containing no negation. A key strategy in proving this decidability is to characterize how a set of states can be reached infinitely often in execution paths of a quantum automaton. For the special case where the quantum automaton has only a single unitary operator and $f$ is an atomic proposition, it is shown based on the Skolem-Mahler-Lech Theorem~\cite{Skolem34, Mahler35, Lech52} that states are reached periodically, and thus the execution can be represented by a cycle graph. In general, we show that this execution graph becomes a general directed graph representing a reversible DFA (deterministic finite automaton), which can be inductively constructed.
\end{itemize}

\subsection{Organization of the paper} The main results are stated in Sec.~\ref{sec:main} after introducing several basic definitions. In Sec.~\ref{sec:Skolem} we first give a brief discussion about the Skolem's problem and relate it to a special case of the quantum reachability problem. Then we prove undecidability of $\G f$, $\U f$ and $\I f$. The undecidability of $\F f$ is separately proved in Sec.~\ref{sec:unde} by using 2-counter Minsky machines. The proofs of decidable results about $\G f$, $\U f$ and $\I f$ for positive $f$ and related algorithms are presented in Sec.~\ref{sec:dec}. A brief conclusion is drawn in Sec.~\ref{sec:con}. Some technical lemmas are collected in Appendix.

\section{Basic Definitions and Main Results}\label{sec:main}
\subsection{A Propositional Logic for Quantum Systems}\label{logic} We first introduce a propositional logical language to describe boolean combinations of the subspaces of a Hilbert space.
Let $\hs$ be the state Hilbert space of a quantum system. A basic property of the system can be described by a (closed) subspace $V$ of $\hs$. In quantum mechanics, to check whether or not this property is satisfied, a binary (yes-no) measurement $\{P_V, P_{V^\perp}\}$ would be performed on the system's current state $\ket{\psi}$, where $P_V$ and $P_{V^\perp}$ are the projection on $V$ and its ortho-complement $V^\perp$, respectively. The measurement outcome is generally nondeterministic: $X$ is considered as being satisfied in $\ket{\psi}$ with probability $\braket{\psi}{P_V}{\psi}$, and it is not satisfied with probability $\braket{\psi}{P_{V^\perp}}{\psi}=1-\braket{\psi}{P_V}{\psi}$.
A quantitative satisfaction relation can be defined by setting a threshold $\lambda\in[0,1]$ to the probability of satisfaction:
$$V\ {\rm is}\ (\lambda,\rhd)-{\rm satisfied\ in}\ \ket{\psi}\ {\rm if}\ \braket{\psi}{P_V}{\psi}\rhd\lambda$$ where $\rhd\in\{<,\leq,>,\geq\}$.
In this paper, we only consider the \textit{qualitative} satisfaction, namely, the $(\lambda,\rhd)-$satisfaction with the threshold $\lambda$ being $0$ or $1$. Obviously, we have:\begin{itemize}
\item $V=\{|\psi\rangle\in\hs|V\ {\rm is}\ (1,\geq)-{\rm satisfied\ in}\ |\psi\rangle\};$
\item $V^\perp=\{|\psi\rangle\in\hs|V\ {\rm is}\ (0,\leq)-{\rm satisfied\ in}\ |\psi\rangle\}.$
\end{itemize}
Thus, it is reasonable to choose the set of atomic propositions to be
$AP=\{V|V\ {\rm is\ a\ (closed)\ subspace\ of}\ \hs\}.$ Furthermore, we define a (classical) propositional logic over $AP$ so that we can talk about, for example, that \textquotedblleft the current state of the quantum system is in subspace $U$, or in $V$ but not in $W$". The logical formulas are generated from $AP$ by using boolean connectives $\neg$, $\wedge$ and $\vee$, and their semantics are inductively defined as follows: for any state $|\psi\rangle\in\hs$,
 \begin{itemize}
\item If $f\in AP$, then $|\psi\rangle\models f$ if $|\psi\rangle\in f$;
\item $|\psi\rangle\models \neg f$ if $|\psi\rangle\models f$ does not hold;
\item $|\psi\rangle\models f_1\wedge f_2$ if $|\psi\rangle\models f_1$ and $|\psi\rangle\models f_2$;
\item $|\psi\rangle\models f_1\vee f_2$ if $|\psi\rangle\models f_1$ or $|\psi\rangle\models f_2$.
\end{itemize} For a logical formula $f$, we write $\|f\|$ for the set of states that satisfy $f$. In general, $\| f\|$ might not be a subspace of $\hs$. For example, for a subspace $V$ of $\hs$, we have:
\begin{itemize}
\item $\|\neg V\|=\{\ket{\psi}\in\hs|V\ {\rm is}\ (1,<)-{\rm satisfied\ in}\ |\psi\rangle\}$;
\item $\|\neg (V^\perp)\|=\{\ket{\psi}\in\hs|V\ {\rm is}\ (0,>)-{\rm satisfied\ in}\ |\psi\rangle\}$.
\end{itemize} It is clear that these classical connectives are different from their quantum counterparts interpreted as the operations in the orthomodular lattice of (closed) subspaces of $\hs$ ~\cite{BN36}.

\subsection{Reachability of Quantum Automata}
\begin{defn} A quantum automaton is a $4-$tuple
$\A=(\hs,Act,\{U_\alpha|\alpha\in Act\},\hs_{ini}),$
where
\begin{enumerate}
\item $\hs$ is the state Hilbert space;
\item $Act$ is a finite set of action names;
\item for each name $\alpha\in Act$, $U_\alpha$ is a unitary operator in $\hs$;
\item $\hs_{ini}\subseteq\hs$ is the subspace of initial states.
\end{enumerate} We say that automaton $\A$ is finite-dimensional if its state space $\hs$ is finite-dimensional. Throughout this paper, we only consider finite-dimensional quantum automata.\end{defn}

A path of $\A$ is generated by successively performing actions, starting in an initial state: $$p=\ket{\psi_0}\overset{U_{\alpha_0}}\rightarrow\ket{\psi_1}\overset{U_{\alpha_1}}\rightarrow\ket{\psi_2}\overset{U_{\alpha_2}}\rightarrow\cdots,$$
where $\ket{\psi_0}\in\hs_{ini}$, $\alpha_n\in Act$, and $\ket{\psi_{n+1}}=U_{\alpha_n}\ket{\psi_n}$
for all $n\geq 0$. For a given initial state $\ket{\psi_0}$ and a sequence of actions $w=\alpha_0\alpha_1\alpha_2\cdots\in {Act}^\omega$, we write the corresponding path as $p=p(\ket{\psi_0},w)$. We further write $\sigma(p)=\ket{\psi_0}\ket{\psi_1}\ket{\psi_2}\cdots$ for the sequence of states in $p$. Sometimes, we simply call $\sigma(p)$ a path of $\A$.

Now let $f$ be a logical formula defined in the above subsection representing a boolean combination of the subspaces of the state Hilbert space, and let
 $\sigma=\ket{\psi_0}\ket{\psi_1}\ket{\psi_2}\cdots$ be an infinite sequence of states in $\hs$. Formally, we define:
\begin{itemize}
\item (Eventually reachable): $\sigma\models \F f$ if $\exists i\geq 0. \ket{\psi_i}\models f$;
\item (Globally reachable): $\sigma\models \G f$ if $\forall i\geq 0. \ket{\psi_i}\models f$;
\item (Ultimately forever reachable): $\sigma\models \U f$ if $\overset{\infty}{\forall} i\geq 0. \ket{\psi_i}\models f$;
\item (Infinitely often reachable): $\sigma\models \I f$ if $\overset{\infty}{\exists} i\geq 0. \ket{\psi_i}\models f$.
\end{itemize}
Here, $\overset{\infty}{\forall}i\geq 0$ means \textquotedblleft $\exists j\geq0, \forall i\geq j$", and $\overset{\infty}{\exists} i\geq 0$ means \textquotedblleft $\forall j\geq 0, \exists i\geq j$".
These reachability properties can be directly applied to quantum automata.

\begin{defn} Let $\A$ be a quantum automaton. Then for $\Delta\in\{\F,\G,\U,\I\}$, we define:
$$\A\models \Delta f\ {\rm if}\ \sigma(p)\models \Delta f\ {\rm for\ all\ paths}\ p\ {\rm in}\ \A.$$
\end{defn}

The reachability of a quantum automaton $\A$ can be stated in a different way. For any action string $s=\alpha_0\alpha_1\cdots\alpha_n\in Act^\ast$, we write $U_s=U_{\alpha_n}\cdots U_{\alpha_1}U_{\alpha_0}.$ If $U_s|\psi_0\rangle\models f$ for some initial state $|\psi_0\rangle\in\hs_{ini}$, then we say that $s$ is accepted by $\A$ with $f$. The set of all accepted action strings is called the language accepted by $\A$ with $f$, and denoted by $\Lan(\A,f)$. We say that a set $S\subseteq Act^\ast$ satisfies the \textit{liveness} property, if
\begin{equation}\label{equ:liveness-condi}
\forall w=\alpha_0\alpha_1\alpha_2\cdots\in Act^\omega,\ \overset{\infty}{\exists} n\geq 0,\ \alpha_0\alpha_1\cdots\alpha_n\in S.
\end{equation}
\begin{lem}\label{lem:relation} Let $\A$ be a quantum automaton with $\dim \hs_{ini}=1$. Then:
\begin{enumerate}
\item $\A\models\F f$ iff $Act^\omega=\Lan(\A,f)\cdot Act^\omega$;
\item $\A\models\I f$ iff $\Lan(\A,f)$ satisfies the liveness condition;
\item $\A\models\G f$ iff $\Lan(\A,f)=Act^\ast$ (i.e. $\Lan(\A,\neg f)=\emptyset$);
\item $\A\models\U f$ iff $Act^\ast-\Lan(\A,f)$ (i.e. $\Lan(\A,\neg f)$) is finite.
\end{enumerate}
Here, $X\cdot Y$ in clause1) is the concatenation of $X$ and $Y$.
\end{lem}
\textit{Proof:} Clauses 1), 2) and 3) can be derived by definition. We only prove clause 4). Let $\ket{\psi_0}\in\hs_{ini}$. If $Act^\ast-\Lan(\A,f)$ is finite, then there exists some $N\geq0$ such that $s\in\Lan(\A,f)$ and thus $U_s\ket{\psi_0}\in \|f\|$ for all action strings $s=\alpha_0\alpha_1\cdots\alpha_n\in Act^\ast$ with $n\geq N$. Furthermore for any path $p$ of $\A$, and $\sigma(p)=\ket{\psi_0}\ket{\psi_1}\cdots$, we have $\ket{\psi_n}\in \|f\|$ for all $n\geq N+1$, and it follows that $\sigma(p)\models\U f$. Therefore, $\A\models\U f$.
On the other hand, if $Act^\ast-\Lan(\A,f)$ is infinite, then according to the K\"{o}nig's infinity lemma, there exists an infinite sequence $w=\alpha_0\alpha_1\cdots\in Act^\omega$ such that
$\overset{\infty}{\exists}n\geq 0,\ \alpha_0\alpha_1\cdots\alpha_n\in Act^\ast-\Lan(\A,f).$
For the corresponding path $p=p(\ket{\psi_0}, w)$, we have $\sigma(p)\not\models\U f$. So $\A\not\models\U f$. $\Box$

\subsection{An Illustrative Example}
\begin{exam}\label{exa:QW}{\rm Consider a quantum walk on a quadrilateral with the state Hilbert space $\hs_4=\spa\{\ket{0},\ket{1},\ket{2},\ket{3}\}$. Its behaviour is described as follows:\begin{enumerate}
\item Initialize the system in state $\ket{0}$.
\item Perform a measurement $\{P_{\rm yes},P_{\rm no}\}$, where $P_{\rm yes}=\op{2}{2},\ P_{\rm no}=I_4-\op{2}{2}.$ Here, $I_4$ is the $4\times 4$ unit matrix. If the outcome is ``yes'', then the walk  terminates; otherwise execute step 3).
\item Nondeterministically choose one of the two unitary operators:
    $$W_{\pm}=\frac{1}{\sqrt{3}}\left(\begin{array}{cccc}
    1 & 1 & 0 & \mp1\\
    \pm1 & \mp1 & \pm1 & 0\\
    0 & 1 & 1 & \pm1\\
    1 & 0 & -1 & \pm1
    \end{array}\right)$$  and apply it.
    Then go to step 2).
\end{enumerate}
It was proved in~\cite{LYY12} that this walk terminates with a probability less than $1$ if and only if a diverging state (i.e. a state with terminating probability $0$) can be reached, and the set of diverging states is $PD_1\cup PD_2$, where\begin{equation*}\begin{split} PD_1&=\spa\{\ket{0},(\ket{1}-\ket{3})/\sqrt{2}\},\\
    PD_2&=\spa\{\ket{0},(\ket{1}+\ket{3})/\sqrt{2}\}.\end{split}\end{equation*}
    So, termination of the walk can be expressed as a reachability property
    $\A\models\G \neg (PD_1\vee PD_2).$ Here, \textquotedblleft$\vee$\textquotedblright\ is obviously boolean disjunction rather than the disjunction in Birkhoff-von Neumann quantum logic.
}\end{exam}

\subsection{Main Theorems}
Now we are ready to present the main problem considered in this paper. For $\Delta\in\{\F,\G,\U,\I\}$, the decision problem for the $\Delta-$reachability is defined as follows:
\begin{prob}\label{prob:q-reach}
Given a finite-dimensional quantum automaton $\A$ and a logical formula $f$ representing a boolean combination of the subspaces of the state Hilbert space of $\A$, decide whether or not $\A\models \Delta f$.
\end{prob}

For the algorithmic purpose, it is reasonable to make the convention: we identify a subspace of $\hs$ with the projection operator on it, and assume that all the projection operators and unitary operators in automaton $\A$ and formula $f$ are represented by complex matrices in a fixed orthonormal basis. Furthermore, we assume that all complex numbers are rational.

The main results of this paper can be stated as the following two theorems:

\begin{thm}\label{the:main-undecidable} (Undecidability) For $\Delta\in\{\F,\G,\U,\I\}$, the problem whether or not $\A\models \Delta f$ is undecidable.
\end{thm}

\begin{thm}\label{the:main-decidable} (Decidability) For $\Delta\in\{\G,\U,\I\}$, if $f$ contains no negation, then the problem whether or not $\A\models \Delta f$ is decidable.
\end{thm}

\section{Relating Quantum Reachability to The Skolem's Problem}\label{sec:Skolem}
\subsection{The Skolem's Problem for Linear Recurrence Sequences}
For convenience of the reader, we first recall several results about the Skolem's problem. A linear recurrence sequence is a sequence $\{a_n\}_{n=0}^\infty$ satisfying a linear recurrence relation given as follows:
\begin{equation}\label{equ:recurrence}
a_{n+d}=c_{d-1}a_{n+d-1}+c_{d-2}a_{n+d-2}+\cdots+c_0a_{n},
\end{equation}
for all $n\geq 0$, where $c_0,c_1,\cdots,c_{d-1}$ are constants with $c_0\neq 0$, and $d$ is called the order of this relation. Let \begin{equation}\label{zero} Z=\{n\in\N|a_n=0\}\end{equation} be the set of indices of null elements of the sequence $\{a_n\}_{n=0}^\infty$. The problem of characterising $Z$ was first studied by Skolem~\cite{Skolem34} in 1934, and his result was generalised by Mahler~\cite{Mahler35} and Lech~\cite{Lech52}.

\begin{thm}[Skolem-Mahler-Lech]\label{the:SML}
In a field of characteristic 0, for any linear recurrence sequence $\{a_n\}_{n=0}^\infty$, the set $Z$ of indices of its null elements is semi-linear; that is, it is the union of a finite set and finitely many arithmetic progressions.
\end{thm}

The above Skolem's problem was further considered in terms of decidability. The problem of deciding whether or not $Z$ is infinite was solved by Berstel and Mignotte~\cite{BM76} who found an algorithm for generating all arithmetic progressions used in Theorem~\ref{the:SML}. The problem of deciding the finiteness of the complement of $Z$ was studied by Salomaa and Soittola~\cite{SS78}. Their results are summarised as the following:

\begin{thm}[Berstel-Mignotte-Salomaa-Soittola]\label{the:skolem-decidable}
For linear recurrence sequences $\{a_n\}_{n=0}^\infty$, it is decidable whether or not
\begin{enumerate}
\item $Z$ is infinite;
\item $Z=\N$;
\item $Z$ contains all except finitely many natural numbers.
\end{enumerate}
\end{thm}

The following emptiness problem dual to item 2) in Theorem~\ref{the:skolem-decidable} was also considered in the literature, but it is still open; for details, we refer to \cite{HHHK05,QW12}.
\begin{prob}\label{prob:empty}
Given a linear recurrence sequence $\{a_n\}_{n=0}^\infty$, decide whether or not $Z$ is empty.
\end{prob}

\subsection{Skolem's Problem in Matrix Form}
In this subsection, we show a useful connection between the quantum reachability problem and the Skolem's problem.
The linear recurrence relation Eq. (\ref{equ:recurrence}) can be written in a matrix form:
\begin{equation}\label{equ:matrix}
a_n=u^TM^nv,
\end{equation}
where $M$ is the $d\times d$ matrix
\[
\left[\begin{array}{ccccc}
c_{d-1}&c_{d-2}&\cdots&c_1&c_0\\
1&0&\cdots&0&0\\
0&1&\cdots&0&0\\
\vdots&\vdots&\vdots&\ddots&\vdots\\
0&0&\cdots&1&0\end{array}\right],
\]
$u=[1,0,\cdots,0]^T$ and $v=[a_{d-1},a_{d-2},\cdots,a_0]^T$ are $d-$dimensional column vectors, and $^T$ stands for transpose. On the other hand, if $\{a_n\}_{n=0}^\infty$ is of form Eq.~(\ref{equ:matrix}) for general $u$,$v$ and $M$ with dimension $d$, then the minimal polynomial $g(x)$ of $M$ is of order at most $d$, $g(M)=0$, and a linear recurrence relation of order no greater than $d$ is satisfied by $\{a_n\}_{n=0}^\infty$. Therefore, the Skolem's problem can be equivalently considered in the matrix form Eq.~(\ref{equ:matrix}).

Let us consider Problem~\ref{prob:q-reach} in a special case: (1) $|Act|=1$, i.e., there is only one unitary operator $U_\alpha$ of $A$, (2) $f=V$ is a subspace of $\hs$, and (3) $\dim\hs_{ini}=\dim V^\perp =1$. Let $\ket{\psi_0}\in \hs_{ini}$ and $|\varphi\rangle\in V^\perp$. Then we have $\Lan(\A,f)=\{n\in\N|\braket{\varphi}{U_\alpha^n}{\psi_0}=0\}.$
It is actually the set $Z$ in Eq. (\ref{zero}) if we think of $U_\alpha$, $\ket{\varphi}$ and $\ket{\psi_0}$ as $M$, $u$, and $v$ in Eq.~(\ref{equ:matrix}). From Lemma~\ref{lem:relation}, the emptiness of $Z$ (Problem~\ref{prob:empty}), and the properties 1), 2) and 3) of $Z$ in Theorem~\ref{the:skolem-decidable} are equivalent to $\A\models\F V$, $\A\models\I V$, $\A\models\G V$, and $\A\models\U V$, respectively. From this point of view, our decidability for a general $f$ (Theorem~\ref{the:main-decidable}) is somewhat a generalization of the decidable results (Theorem~\ref{the:skolem-decidable}) of Skolem's problem where $f$ is taken to be a subspace.

\subsection{Undecidability of $\A\models\G f$, $\A\models\U f$ and $\A\models\I f$}
Now we consider an undecidable result relevant to the Skolem's problem. Instead of $\{M^n|n\in\N\}$ in Eq.~(\ref{equ:matrix}), there is a semi-group generated by a finite number of matrices $M_1,M_2,\cdots,M_k$, written as $\<M_1,M_2,\cdots,M_k\>$. Then the emptiness problem can be generalised as follows:
\begin{prob}\label{prob:gener-empty}
Provided $d\times d$ matrices $M_1,M_2,\cdots,M_k$ and $d$-dimensional vectors $u$ and $v$, decide whether or not $\exists M\in \<M_1,M_2,\cdots,M_k\>$ s.t. $u^TMv=0$.
\end{prob}

The above problem was proved to be undecidable in \cite{Paz71} and \cite{CK98}, through a reduction from the Post's Correspondence Problem (PCP)~\cite{Pos46}. Similar to the discussion in last subsection, we can choose $M_i$ as unitary operators and $u$, $v$ as quantum states, and then the emptiness of $\Lan(\A,f)$ for $f=V$ and $\dim\hs_{ini}=\dim V^\perp =1$ but with $|Act|>1$ being allowed can be regarded as a special case of Problem \ref{prob:gener-empty}. In fact, this problem was also proved to be undecidable by Blondel et. al.~\cite{BJKP05}.
\begin{thm}[Blondel-Jeandel-Koiran-Portier]\label{the:unitary-undeci}
It is undecidable whether or not $\Lan(\A,V)$ is empty, given a quantum automaton $\A$ and a subspace $V$ with $\dim\hs_{ini}=\dim V^\perp =1$.
\end{thm}
\smallskip

We can use this undecidable result to prove the Theorem~\ref{the:main-undecidable} for $\Delta\in\{\G,\U,\I\}$. We first prove undecidability of $\A\models\G f$. Let automaton $\A$ be the same as in Theorem~\ref{the:unitary-undeci}, but put $f=\neg V$ (not $V$). Then according to clause 3) of Lemma~\ref{lem:relation}, $\A\models\G f$ is equivalent to the emptiness of $\Lan(\A,\neg(\neg V))=\Lan(\A,V)$. The undecidability follows immediately from Theorem~\ref{the:unitary-undeci}.

To prove undecidability of $\A\models\U f$ and $\A\models\I f$, we slightly modify each quantum automaton $\A=(\hs,Act,\{U_\alpha|\alpha\in Act\},\hs_{ini})$ by adding a silent action $\tau$. Assume that $\tau\notin Act$ and $U_\tau=I$ (the identity operator in $\hs$). Put $\A^\prime=(\hs,Act\cup\{\tau\},\{U_\alpha|\alpha\in Act\cup\{\tau\}\},\hs_{ini}).$ Then we claim:
\begin{equation}\label{xxx}\A\models\G f\ {\rm iff}\ \A^\prime\models\U f\ {\rm iff}\ \A^\prime\models\I f.\end{equation}
In fact, it is obvious that $\A\models\G f\Rightarrow \A^\prime\models\U f\Rightarrow \A^\prime\models\I f$ because $U_\tau$ is silent. Conversely, if $\A\not\models\G f$, then there exists $s=\alpha_0\alpha_1\cdots\alpha_n\in Act^*$ such that
$U_s\ket{\psi_0}\not\models f$. We consider the infinite sequence of actions $w=s\tau^\omega\in (Act\cup\{\tau\})^\omega$. It is clear that
$\sigma(p(\ket{\psi_0},w))\not\models\U f$ and $\sigma(p(\ket{\psi_0},w))\not\models\I f$, and so $\A'\not\models\U f$ and $\A'\not\models\I f$. Finally, undecidability of $\A\models\U f$ and $\A\models\I f$ follows immediately from Eq.~(\ref{xxx}) and undecidability of $\A\models\G f$. Remarkably, the simple form of $f=\neg V$ is sufficient for undecidability.

\section{Reduction from The halting problem for 2-counter Minsky machines}\label{sec:unde}

The aim of this section is to prove undecidability of $\A\models \F f$. Our strategy is a reduction from the halting problem for 2-counter Minsky machines to reachability of quantum automata.

\subsection{2-counter Minsky Machine}

A 2-counter Minsky machine~\cite{Min76} is a program $\M$ consisting of two variables (counters) $a$ and $b$ of natural numbers $\N$, and a finite set of instructions, labeled by $l_0,l_1,\cdots,l_m$. This program starts at $l_0$ and halts at $l_m$. Each of instructions $l_0,l_1,\cdots,l_{m-1}$ is one of the following two types:\\
\begin{tabular}{rll}
\textbf{increment}&\ \ \ \ $l_i:$&$c\leftarrow c+1;$ goto $l_j;$\\
\textbf{test-and-decrement}&\ \ \ \ $l_i:$&if $c=0$ then goto $l_{j_1};$\\
&&else $c\leftarrow c-1;$ goto $l_{j_2};$
\end{tabular}
\smallskip
\\where $c\in\{a,b\}$ is one of the counters. The halting problem is as follows: given a 2-counter Minsky machine $\M$ together with the initial values of $a$ and $b$, decide whether the computation of $\M$ will terminate or not. This problem is known to be undecidable.

For convenience of relating $\M$ to a quantum automaton, we slightly modify the definition of $\M$ without changing its termination:
\begin{enumerate}
\item Without loss of generality, we assume the initial values of $a$ and $b$ to be both $0$. This can be done because any value can be achieved from zero by adding some instructions of increment at the beginning.
\item For each instruction $l_i$ of test-and-decrement of $c$, we rewrite it as
    \begin{equation}\label{equ:Lprime}\begin{split}
    l_i:&\ \ {\rm if}\ c=0\ {\rm then\ goto}\ l_i';\ {\rm else\ goto}\ l_i^{\prime\prime};\\
    l_i':&\ \ {\rm goto}\ l_{j_1};\\
    l_i^{\prime\prime}:&\ \ c\leftarrow c-1;\ {\rm goto}\ l_{j_2};
    \end{split}\end{equation}
    where $l_i'$ and $l_i{''}$ are new instructions. For $c\in\{a,b\}$, we write $L_{1c}$ for the set of all instructions of increment of $c$; and we write $L_{2c}$, $L_{2c}'$ and $L_{2c}^{\prime\prime}$ for the set of instructions $l_i$, the set of instructions $l_i'$ and the set of instructions $l_i^{\prime\prime}$ given in Eq.~(\ref{equ:Lprime}), respectively. Now the set of all instructions of $\M$ becomes
    $$L=L_{1a}\cup L_{1b}\cup L_{2a}\cup L_{2b}\cup L_{2a}'\cup L_{2b}'\cup L_{2a}^{\prime\prime}\cup L_{2b}^{\prime\prime}\cup\{l_m\}.$$
\item We rewrite $l_m$ as $$l_m:\ \ {\rm goto}\ l_m;$$ and we define that $\M$ terminates if $l_m$ is reachable during the computation.
\end{enumerate} Obviously, the halting problem is also undecidable for 2-counter Minsky machines defined in this way.

We will encode 2-counter Minsky machines into quantum automata so that undecidability of $\A\models\F f$ is derived from the undecidability of halting problem. More precisely, for any given 2-counter Minsky machine $\M$, we will construct a quantum automaton $\A$ and find two subspaces $V$ and $W$ of $\hs$ such that
\begin{equation}\label{equ:eqv}\M\ {\rm terminates} \Leftrightarrow \A\models\F (V\wedge \neg W).\end{equation} The basic ideas of this construction are outlined as follows: \begin{enumerate}
\item A state of $\M$ is of form $(a,b,x)$, where $a,b\in\N$ are the values of the two counters, and $x\in L$ is the instruction to be executed immediately. We will use quantum states $\ket{\phi_n}$ and $\ket{l}$ to encode nature numbers $n$ and instructions $l$, respectively. Then the corresponding quantum state in $\A$ is chosen as the product state $\ket{\psi}=\ket{\phi_a}\ket{\phi_b}\ket{l}$.
\item The computation of $\M$ is represented by the sequence of its states: \begin{equation}\label{equ:sigmaM}\sigma_\M=(a_0,b_0,x_0)(a_1,b_1,x_1)(a_2,b_2,x_2)\cdots,\end{equation}
    where $(a_0,b_0,x_0)=(0,0,l_0)$ is the initial state and $(a_{i+1},b_{i+1},x_{i+1})$ is the successor of $(a_i,b_i,x_i)$ for all $i\geq 0$. We will construct unitary operators of $\A$ to encode the transitions from a state to its successor. Then by successively taking the corresponding unitary operators, the quantum computation\begin{equation}\label{equ:sigma0}\sigma_0=\ket{\psi_0}\ket{\psi_1}\cdots,\ \forall i\geq 0\ \ket{\psi_i}=\ket{\phi_{a_i}}\ket{\phi_{b_i}}\ket{x_i}\end{equation}
    is achieved in $\A$ to encode $\sigma_\M$.
\item From the correspondence between $\sigma_\M$ and $\sigma_0$, termination of $\M$ will be encoded as certain reachability property of $\sigma_0$ (Lemma~\ref{lem:sigma-in}).
\item Besides $\sigma_0$, infinitely many computation paths are achievable in $\A$. So there is still a gap between reachability of $\sigma_0$ and that of $\A$. Our solution is to construct two subspaces $V$ and $W$ such that $\sigma\models\F (V\wedge\neg W)$ for all paths $\sigma$ of $\A$ except $\sigma_0$ (Lemma~\ref{lem:other-sigma}). Then $$\A\models \F (V\wedge \neg W)\Leftrightarrow\sigma_0\models \F (V\wedge \neg W),$$ and Eq.~(\ref{equ:eqv}) will be proved from this equivalence.
\end{enumerate}

\subsection{Encoding Classical States into Quantum States}
This subsection is the first step of constructing quantum automaton $\A$. We show how to encode the states of $\M$ into quantum states in a finite dimensional Hilbert space. First, we use qubit states in the $2-$dimensional Hilbert space $\hs_2=\spa\{\ket{0},\ket{1}\}$ to encode natural numbers. Consider the following unitary operator acting on $\hs_2$:
$$G=\op{+}{+}+\me^{\mi\theta}\op{-}{-},$$
where $\ket{\pm}=(\ket{0}\pm\ket{1})/\sqrt{2}$ and $\me^{\mi\theta}=(3+4i)/5$. It is easy to see that for any integer $n$, $G^n\ket{0}=\ket{0}\Leftrightarrow n=0.$ So for each integer $n$, we can use $G^n\ket{0}$ to encode $n$. Moreover, operator $G$ can be thought of as the successor function $g(n)=n+1$.
Now, let $\hs_a=\hs_b=\hs_2$ and we use states in $\hs_a$ and $\hs_b$ to encode the counters $a$ and $b$, respectively. Specifically, for each value $n$ of $c\in\{a,b\}$, the corresponding state is $\ket{\phi_n}=G_c^n\ket{0}\in\hs_c$.

We simply encode the instruction labels $l$ as orthonormal quantum states $\ket{l}$ and construct the Hilbert space $\hs_L=\spa\{\ket{l}|l\in L\}$. Then a state $(a,b,x)$ of $\M$ can be encoded as the quantum state $\ket{\phi_a}\ket{\phi_b}\ket{x}\in\hs_a\otimes\hs_b\otimes\hs_L.$
Moreover, the computation $\sigma_\M$ of $\M$ is encoded as the sequence $\sigma_0$ of quantum states.
We note that $\M$ terminates if and only if $x_i=l_m$ for some state $(a_i,b_i,x_i)$ in $\sigma_\M$. This condition is equivalent to $\ket{\psi_i}\in V_0$, where
\begin{equation}\label{equ:V0}V_0=\hs_a\otimes\hs_b\otimes\spa\{\ket{l_m}\}.\end{equation} So the termination of $\M$ is reduced to reachability of $\sigma_0$ as follows:
\begin{lem}\label{lem:sigma-in} $\M$ terminates iff $\sigma_0\models\F V_0$.\end{lem}

\subsection{Construction of Unitary Operators of $\A$}
In this subsection, we construct unitary operators of $\A$ to encode the state transitions of $\M$. For any state $(a,b,x)$ of $\M$, we consider the transition from this state to its successor. There are two cases:\begin{enumerate}
\item $x\in L_{1a}\cup L_{1b}\cup L_{2a}'\cup L_{2b}'\cup L_{2a}^{\prime\prime}\cup L_{2b}^{\prime\prime}\cup\{l_m\}$. Then from the definition of $L$, $x$ is of form
    $$x:\ \ c\leftarrow c+e;\ {\rm goto}\ y;$$
    where $c\in\{a,b\}$, $y\in L$ and $e=1,0,-1$ for $l\in L_{1c}$, $L_{2c}'\cup\{l_n\}$, $L_{2c}^{\prime\prime}$, respectively. So the successor of $(a,b,x)$ is as $(\tilde{a},\tilde{b},y)$, where $\tilde{a}=a+e$, $\tilde{b}=b$ for $c=a$, and $\tilde{a}=a$, $\tilde{b}=b+e$ for $c=b$.
    We construct a unitary operator corresponding to $x$:
    $$U_x=O_c^e\otimes O_{xy},$$
    where $O_a=G_a\otimes I_b$ and $O_b=I_a\otimes G_b$ are unitary operators on $\hs_a\otimes\hs_b$, and $O_{xy}$ is a unitary operator on $\hs_{L}$ satisfying $O_{xy}\ket{x}=\ket{y}$. Obviously, we have $\ket{\phi_{\tilde{a}}}\ket{\phi_{\tilde{b}}}\ket{y}=U_x\ket{\phi_a}\ket{\phi_b}\ket{x}$ for any $a$, $b$. So $U_x$ is what we want.
\item $x\in L_{2a}\cup L_{2b}$. Then $x$ is of form
    $$x:\ \ {\rm if}\ c=0\ {\rm then\ goto}\ y;\ {\rm else\ goto}\ z;$$
    where $c\in\{a,b\}$, $y\in L_{2c}'$ and $z\in L_{2c}^{\prime\prime}$. The successor of $(a,b,x)$ is $(a,b,y)$ for $c=0$, and is $(a,b,z)$ for $c\neq 0$. We construct two unitary operators corresponding to $x$: $$U_{x0}=I_a\otimes I_b\otimes O_{xy}\ {\rm and}\ U_{x1}=I_a\otimes I_b\otimes O_{xz},$$ where $O_{xy}\ket{x}=\ket{y}$ and $O_{xz}\ket{x}=\ket{z}$. Thus, $U_{x0}$ is used when $c=0$, and $U_{x1}$ is used when $c\neq 0$.
    \end{enumerate}

Now, we only need to specifically construct the unitary operator $O_{xy}$ for given $x,y\in L$. To this end, we construct for each $l\in L$ a new quantum state $\ket{\hat{l}}$ to be the result of $O_{xy}\ket{l}$ (for $x\neq l$). Formally, we construct a new state space $\hat{\hs}_L=\spa\{\ket{\hat{l}}:x\in L\}$ and extend $\hs_L$ to $$\hs_{2L}=\hs_L\oplus\hat{\hs}_L=\spa\{\ket{l},\ket{\hat{l}}|l\in L\}.$$
Then $O_{xy}$ is defined in $\hs_{2L}$ as
\begin{equation}\label{zzz}\begin{split}
    O_{xy}\ket{x}&=\ket{y},\ O_{xy}\ket{l}=\ket{\hat{l}}\ (\forall l\in L,l\neq x),\\
    O_{xy}\ket{\hat{y}}&=\ket{\hat{x}},\ O_{xy}\ket{\hat{l}}=\ket{l}\ (\forall l\in L,l\neq y).
    \end{split}\end{equation}
Notably, $O_{xy}$ satisfies the following property:
\begin{equation}\label{equ:img}
O_{xy}\ket{z}\in\hat{\hs}_L,\forall z\in L\ {\rm and}\ z\neq x.
\end{equation}

Finally, quantum automaton $\A$ is constructed as follows: the state space is $\hs=\hs_a\otimes\hs_b\otimes\hs_{2L}$, the unitary operators are $\{U_\alpha|\alpha\in Act\}$, where
$$Act=\{x0,x1| x\in L_{2a}\cup L_{2b}\}\cup L\backslash(L_{2a}\cup L_{2b}),$$
and the initial state is $\ket{\psi_0}=\ket{0}\ket{0}\ket{l_0}$. From the construction of the unitary operators, we see that the sequence $\sigma_0$ of quantum states defined by Eq.~(\ref{equ:sigma0}) is achievable in $\A$.

\subsection{Construction of $V$ and $W$}
This subsection is the last step to achieve Eq.~(\ref{equ:eqv}): construction of subspaces $V$ and $W$. First, we find a way to distinguish $\sigma_0$ from other paths of $\A$. Specifically, we consider a state $\ket{\psi_n}=\ket{\phi_{a_n}}\ket{\phi_{b_n}}\ket{x_n}$ in $\sigma_0$ to be transformed by a ``mismatched'' unitary operator in $\{U_\alpha|\alpha\in Act\}$; namely, this unitary operator transforms $\ket{\psi_n}$ into a state $\ket{\psi'}$ other than $\ket{\psi_{n+1}}$. Each unitary operator in $\A$ is of form $U_y$, $U_{y0}$, or $U_{y1}$, where $y$ is the corresponding instruction. If $y\neq x_n$, then it is definitely mismatched. It follows from the Eq.~(\ref{equ:img}) that $\ket{\psi'}\in \hat{V}$, where $\hat{V}=\hs_a\otimes\hs_b\otimes\hat{\hs}_{L}.$

Now we only need to consider the case of $y=x_n$. We have $x_n\in L_{2a}\cup L_{2b}$, because there are two unitary operators corresponding to $x_n$: the one mismatched and the one not. For $x\in L_{2a}$, there are two cases:
\begin{enumerate}
\item $a_n=0$ and the mismatched unitary operator is $U_{x_n1}$. From the definition of $U_{x_n1}$, we have $$\ket{\psi'}=U_{x_n1}\ket{0}\ket{\phi_{b_n}}\ket{x_n}=\ket{0}\ket{\phi_{b_n}}\ket{z},$$ where $z\in L_{2a}^{\prime\prime}$. We write $$V_{2a}=\spa\{\ket{0}\}\otimes\hs_b\otimes\spa\{\ket{l}:l\in L_{2a}^{\prime\prime}\}.$$ Then $\ket{\psi'}\in V_{2a}$.
\item $a_n>0$ and the mismatched one is $U_{x_n0}$. From the definition of $U_{x_n0}$, we have
    $$\ket{\psi'}=U_{x_n0}\ket{\phi_{a_n}}\ket{\phi_{b_n}}\ket{x_n}=\ket{\phi_{a_n}}\ket{\phi_{b_n}}\ket{y},$$
    where $y\in L_{2a}'$. We write\begin{equation*}\begin{split} V_{1a}&=\hs_a\otimes\hs_b\otimes\spa\{\ket{l}:l\in L_{2a}'\},\\ W_a&=\spa\{\ket{0}\}\otimes\hs_b\otimes\spa\{\ket{l}:l\in L_{2a}'\}.\end{split}\end{equation*} Then $\ket{\psi'}\in V_{1a}\backslash W_a$.
\end{enumerate}
Similarly, for $x_n\in L_{2b}$ we can prove that $\ket{\psi'}\in V_{2b}$ for $b_n=0$ and $\ket{\psi'}\in V_{1b}\backslash W_b$ for $b_n>0$, where\begin{equation*}\begin{split}
V_{1b}&=\hs_a\otimes\hs_b\otimes\spa\{\ket{l}:l\in L_{2b}'\},\\
V_{2b}&=\hs_a\otimes\spa\{\ket{0}\}\otimes\spa\{\ket{l}:l\in L_{2b}^{\prime\prime}\},\\
W_b&=\hs_a\otimes\spa\{\ket{0}\}\otimes\spa\{\ket{l}:l\in L_{2b}'\}.
\end{split}\end{equation*}

We have actually proved that a state
\begin{equation}\label{equ:in5}
\ket{\psi'}\in \hat{V}\cup(V_{1a}\backslash W_a)\cup(V_{1b}\backslash W_b)\cup V_{2a}\cup V_{2b}\end{equation} is always reachable in computation paths of $\A$ other than $\sigma_0$. On the other hand, it is also easy to verify that such a state cannot be in $\sigma_0$. So $\sigma_0$ can be distinguished by this reachability property.

Now we put\begin{equation*}\begin{split}
V&=V_0+\hat{V}+V_{1a}+V_{1b}+V_{2a}+V_{2b},\\
W&=W_a+W_b,
\end{split}\end{equation*} where $V_0$ is defined by Eq.~(\ref{equ:V0}). Then we have:
\begin{lem}\label{lem:other-sigma} For all paths $p$ in $\A$ with state sequences $\sigma(p)\neq\sigma_0$, we have $\sigma(p)\models\F (V\wedge \neg W)$.\end{lem}
\textit{Proof:} We only need to note that the union of five sets in Eq.~(\ref{equ:in5}) is included in $\{0\}\cup(V\backslash W)$, and then this result is straightforward from our discussion above. $\Box$
\smallskip

Moreover, we have the following result:
\begin{lem}\label{lem:sigma-satisfy}$\sigma_0\models\F (V\wedge \neg W)$ iff $\sigma_0\models\F V_0$.\end{lem}
\textit{Proof:} It suffices to prove that for any state $\ket{\psi_n}$ in $\sigma_0$,
$$\label{equ:VW-V0}\ket{\psi_n}\in V\backslash W\ {\rm iff}\ \ket{\psi_n}\in V_0.$$
The ``if'' part is obvious since $V_0\subseteq V$ and $V_0\cap W=\{0\}$. We now prove the ``only if'' part. As $\ket{\psi_n}=\ket{\phi_{a_n}}\ket{\phi_{b_n}}\ket{x_n}$ is a state in $\sigma_0$, $(a_n,b_n,x_n)$ is a state in $\sigma_\M$ and thus $x_n\in L$. From the definition of $L$ and Eq.~(\ref{equ:Lprime}), $\ket{\psi_n}$ is checked in the following cases of $x_n$:
\begin{equation*}\begin{split}
&x_n\in L_{1a}\cup L_{1b}\cup L_{2a}\cup L_{2b},\ {\rm thus}\ \ket{\psi_n}\not\in V;\\
&x_n\in L_{1a}'\Rightarrow a_n=0,\ {\rm thus}\ \ket{\psi_n}\in W_a;\\
&x_n\in L_{1b}'\Rightarrow b_n=0,\ {\rm thus}\ \ket{\psi_n}\in W_b;\\
&x_n\in L_{2a}^{\prime\prime}\Rightarrow a_n\neq 0,\ {\rm thus}\ \ket{\psi_n}\notin V;\\
&x_n\in L_{2b}^{\prime\prime}\Rightarrow b_n\neq 0,\ {\rm thus}\ \ket{\psi_n}\notin V.
\end{split}\end{equation*}
None of them satisfies $\ket{\psi_n}\in V\backslash W$. So the only possibility is $x_n=l_m$, and then $\ket{\psi_n}\in V_0$. $\Box$

Finally, we obtain Eq.~(\ref{equ:eqv}) by simply combining Lemmas \ref{lem:sigma-in}, ~\ref{lem:other-sigma} and~\ref{lem:sigma-satisfy}. Undecidability of $\A\models\F f$ is so proved, even for the simple form of $f=V\wedge\neg W$.

\section{Decidable Results}\label{sec:dec}
We prove Theorem~\ref{the:main-decidable} in this section. We write $f$ in the disjunctive normal form.
As it contains no negation, for each conjunctive clause $f_i$ of $f$, $\|f_i\|$ is a subspace of $\hs$. We write $V_i=\|f_i\|\in AP$, then $f$ can be equivalently written as $f=\bigvee_{i=1}^m V_i$ and $\|f\|=\bigcup_{i=1}^m V_i$ is a union of finitely many subspaces of the state Hilbert space $\hs$ of quantum automaton $\A$.

To decide whether or not $\A\models\Delta f$, we need to compute the set of all predecessor states with respect to a reachability property. Formally, for any given quantum automaton $\A=(\hs,Act,\{U_\alpha|\alpha\in Act\},\hs_{ini})$ and any $\ket{\psi}\in\hs$, we consider the automaton $\A(\psi)=(\hs,Act,\{U_\alpha|\alpha\in Act\},\spa\{\ket{\psi}\})$ for the paths starting in $\ket{\psi}$. Then for any $\Delta\in\{\G,\U,\I\}$, $\ket{\psi}$ is called a $(\Delta,f)-$predecessor state if $\A(\psi)\models\Delta f$, and we write the set of all predecessor states as
$$Y(\A,\Delta,f)=\{\ket{\psi}\in\hs|\A(\psi)\models\Delta f\}.$$
Then $\A\models\Delta f$ can be decided by checking whether or not $\hs_{ini}\subseteq Y(\A,\Delta,f)$.

\subsection{Decidability of $\A\models\I f$ for Single Unitary Operator}
We will prove the decidability of $\A\models\I f$ by constructing the set $Y(\A,\I,f)$. In this subsection, we do this for a special case in which $|Act|=1$ and $m=1$, i.e., $\A$ contains only a single unitary operator, and $f=V$ is a subspace. It should be pointed out that the result for this special case was proved in \cite{BM76} as the decidability of finiteness Skolem's problem in the single matrix form. Here, we present our new proof as it would be useful for us to obtain a general result for finitely many unitary operators in next subsection. For convenience, we simply write $Y$ for $Y(\A,\I,f)$ in these two subsections.

Let $Act=\{\alpha\}$, and the string $\alpha^n$ is simply represented by $n$. By an algorithm, we show that $Y$ is a union of finitely many subspaces $Y_0,Y_1,\cdots,Y_{p-1}$ which forms a cycle graph under the unitary transformation, namely $Y_{r+1}=U_\alpha Y_r$ for all $0\leq r\leq p-2$ and $Y_0=U_\alpha Y_{p-1}$. Then $Y$ can be written as $Y=\bigcup_{r=0}^{p-1}U_\alpha^rY_0$ and $Y_0=U_\alpha^pY_0$. The following lemma is required for proving correctness of our algorithm.

\begin{lem}\label{lem:period-p}
For any unitary operator $U$ on $\hs$, there exists a positive integer $p$ such that for any subspace $K$ of $\hs$, $U^pK=K$ provided $U^nK=K$ for some integer $n$. We call this integer $p$ the period of $U$.
\end{lem}

We put the technical proof of the above lemma into Appendix A. Now $Y$ can be computed by Algorithm \ref{alg:check1}.
\begin{algorithm}\begin{enumerate}
\item Compute the period $p$ of $U_\alpha$;
\item Compute the maximal subspace $K$ of $V$ such that $U_\alpha^pK=K$;
\item $Y=\bigcup_{r=0}^{p-1}U_\alpha^rK.$
\end{enumerate}\caption{}\label{alg:check1}
\end{algorithm}
Step 1) can be done as described in the proof of Lemma~\ref{lem:period-p}. Step 2) can be done as follows: initially put $K_0=V$, repeatedly compute $K_{n+1}=K_n\cap U_\alpha^pK_n$ until $K_{n+1}=K_{n}$, and then $K=K_n$. Sometimes, we write $K$ as $K(U_\alpha,V)$ to show dependence of $K$ on $U_\alpha$ and $V$. Correctness of this algorithm is proved in Appendix B.

\subsection{Decidability of $\A\models\I f$ for General Case}
Now, we construct $Y=Y(\A,\I,f)$ for a general input: $\A$ and $f=\bigvee_{i=1}^m V_i$. Like the case of single unitary operator, we can prove that $Y$ is a union of finitely many subspaces. The result can be specifically described as follows:
\begin{lem}\label{lem:3-condi}
Let $X=\{Y_1,Y_2,\cdots,Y_q\}$ be a set of subspaces of $\hs$ satisfying the following three conditions:
\begin{enumerate}
\item For any $Y_i$ and $\alpha\in Act$, there exists $Y_j$ such that $U_\alpha Y_i=Y_j$. In other words, under the unitary transformations, these subspaces form a more general directed graph than a simple cycle graph in the case of single unitary operator.
\item For any simple loop (namely $Y_{r_i}\neq Y_{r_j}$ for different $i$ and $j$ in the loop) $$Y_{r_0}\overset{U_{\alpha_0}}\rightarrow Y_{r_1}\overset{U_{\alpha_1}}\rightarrow \cdots\overset{U_{\alpha_{k-2}}}\rightarrow Y_{r_{k-1}}\overset{U_{\alpha_{k-1}}}\rightarrow Y_{r_0},$$
there exists some $i\in\{0,1,\cdots,k-1\}$ and $j\in\{1,2,\cdots,m\}$ such that $Y_{r_i}\subseteq V_j$.
\item $Y\subseteq Y_1\cup Y_2\cup\cdots\cup Y_q$.
\end{enumerate}
Then $Y=Y_1\cup Y_2\cup\cdots\cup Y_q$.
\end{lem}

\textit{Proof:} From condition 3), it suffices to prove that if $X$ satisfies the first two conditions, then $\cup X\subseteq Y$. We only need to prove that for any $\ket{\psi_0}\in \cup X$, we have $\ket{\psi_0}\in Y$, namely, $\A(\psi_0)\models\I f$. From the definition, it suffices to prove that $$\forall w=\alpha_0\alpha_1\cdots\in {Act}^\omega\ \overset{\infty}{\exists}n\geq 0\ {\rm s.t.}\ \ket{\psi_n}\in \|f\|,$$  where $\ket{\psi_{n+1}}=U_{\alpha_n}\ket{\psi_n}$ for $n=0,1,\cdots$.

Now we choose $Y_{r_0}\in X$ such that $\ket{\psi_0}\in Y_{r_0}$. According to the first condition, let $Y_{r_{n+1}}=U_{\alpha_n}Y_{r_n}$, $n=0,1,\cdots$. Then $\ket{\psi_n}\in Y_{r_n}$. Consider any pairs of $r_i$ and $r_j$ such that $i<j$, $r_i=r_j$, and $r_i,r_{i+1},\cdots,r_{j-1}$ are pairwise different. Applying the second condition in the simple loop
$$Y_{r_i}\overset{U_{\alpha_i}}\rightarrow Y_{r_{i+1}}\overset{U_{\alpha_{i+1}}}\rightarrow \cdots\overset{U_{\alpha_{j-2}}}\rightarrow Y_{r_{j-1}}\overset{U_{\alpha_{j-1}}}\rightarrow Y_{r_i},$$ there exists some $n$ such that $i\leq n<j$ and $Y_{r_{n}}\subseteq \|f\|$. Then $\ket{\psi_n}\in \|f\|$. As we can choose infinitely many pairs $(r_i,r_j)$ in the sequence $w$, we can find infinitely many $n$'s. Thus $\ket{\psi_0}\in Y$. $\Box$

Therefore, to construct $Y$ we only need to find an algorithm for constructing a set of subspaces $X=\{Y_1,Y_2,\cdots,Y_q\}$ satisfying the three conditions of Lemma~\ref{lem:3-condi}. To this end, we invoke a lemma which is proved in \cite{LYY12}:

\begin{lem}\label{lem:chain}
Suppose that $X_k$ is the union of a finite number of subspaces of
$\hs$ for all $k\geq 0$. If
$X_0\supseteq X_1\supseteq\cdots\supseteq X_k\supseteq\cdots,$
then there exists $n\geq 0$ such that $X_k=X_n$ for all $k\geq n$.
\end{lem}

Now the set $X$ can be computed by Algorithm \ref{alg:check2}.
\begin{algorithm}\begin{enumerate}
\item Initially put $X\leftarrow\{\hs\}$ then jump to step 2);
\item If $X$ satisfies condition 1) and condition 2) of Lemma~\ref{lem:3-condi}, then return $X$; otherwise construct a new set $X'$ of subspaces of $\hs$ satisfying
     $Y\subseteq \cup X'\subset\cup X$, and put $X\leftarrow X'$, then repeat step 2). Here notation ``$\subset$" is for ``proper subset".
\end{enumerate}\caption{}\label{alg:check2}
\end{algorithm}
Step 2) is the key step in the algorithm, in which $X$ can be replaced by a ``smaller" one $X'$ if it is not available. Due to Lemma~\ref{lem:chain}, this step can only be executed a finite number of times and thus an output $X$ satisfying condition 1) and condition 2) of Lemma \ref{lem:3-condi} should be returned by the algorithm. We also note that condition 3) of Lemma \ref{lem:3-condi} is always satisfied by $X$ during the execution. So this output is just what we need.

Now we give a detailed description of step 2). It can be properly formalized as a lemma:
\begin{lem}\label{lem:proper-W}
Given a set $X=\{Y_1,Y_2,\cdots,Y_q\}$ of subspaces in which any two subspaces $Y_i$ and $Y_j$ do not include each other, if $X$ satisfies condition 3) but does not satisfy condition 1) or condition 2) of Lemma \ref{lem:3-condi}, then we can algorithmically find some $Y_i\in X$ and its proper subspaces $W_1,W_2,\cdots,W_l$, such that
\begin{equation}\label{equ:included}
Y\cap Y_i\subseteq W_1\cup W_2\cup\cdots\cup W_l.\end{equation}
\end{lem}

The proof of the above lemma is postponed to Appendix C. From this lemma, we can construct $X'$ for any given $X$ as follows. First, we eliminate all such $Y_i$ from $X$ that $Y_i\subset Y_j$ for some $Y_j\in X$. Then from Lemma~\ref{lem:proper-W} we can find some $Y_i\in X$ and its subspaces $W_1,W_2,\cdots,W_l$ satisfying Eq.~(\ref{equ:included}).
We put
$X'=X\cup\{W_k|1\leq k\leq l\}\backslash\{Y_i\},$
and then $\cup X'\subset\cup X$. As $Y\subseteq \cup X$, we also have $Y\subseteq\cup X'$ from Eq.~(\ref{equ:included}).

\subsection{Decidability of $\A\models\G f$ and $\A\models\U f$}
We now prove Theorem~\ref{the:main-decidable} for $\Delta\in\{\G,\U\}$. We first prove the decidability of $\A\models\G f$ by computing $Y=Y(\A,\G,f)$. According to clause 3) in Lemma~\ref{lem:relation}, we have
\begin{equation}\begin{split}
Y&=\{\ket{\psi}\in\hs|\Lan(\A(\psi),f)=Act^\ast\}\\
&=\{\ket{\psi}\in\hs|U_s\ket{\psi}\in\|f\|,\forall s\in Act^\ast\}.
\end{split}\end{equation}
Then we obtain $\forall \alpha\in Act,\ U_\alpha Y\subseteq Y\subseteq\|f\|$. In fact, $Y$ can be computed by Algorithm~\ref{alg:check3}, and thus $Y$ is the maximal one of sets satisfying $\forall \alpha\in Act,\ U_\alpha Y= Y\subseteq\|f\|$.
\begin{algorithm}\begin{enumerate}
\item $Y\leftarrow V_1\cup V_2\cup\cdots\cup V_m$;
\item If $U_\alpha Y\neq Y$, for some $\alpha\in Act$, then $Y\leftarrow U_\alpha^{-1}Y\cap Y;$ otherwise return $Y$.
\end{enumerate}\caption{}\label{alg:check3}
\end{algorithm}

\textit{Correctness of Algorithm \ref{alg:check3}:} We write $Y_0,Y_1,\cdots$ for the instances of $Y$ during the execution of the algorithm. Then $Y_0=V_1\cup V_2\cup\cdots\cup V_m$ and $Y_{n+1}=U_\alpha^{-1}Y_n\cap Y_n$ for some $\alpha\in Act$. It can be proved by induction on $n$ that each $Y_n$ is a union of finitely many subspaces of $\hs$. Note that $Y_0\supset Y_1\supset Y_2\supset\cdots$
is a descending chain. According to Lemma~\ref{lem:chain}, this chain would terminates at some $n$, and the algorithm output is $Y_n$. We have $U_\alpha Y_n=Y_n$ for all $\alpha\in Act$. Now we prove $Y_n=Y$. First, since $Y\subseteq\|f\|=Y_0$ and $Y\subseteq U_\alpha^{-1}Y$ for all $\alpha\in Act$, it can be proved by induction on $k$ that $Y\subseteq Y_k$ for all $k$, and particularly, $Y\subseteq Y_n$. On the other hand, As $U_s Y_n=Y_n\subseteq \|f\|$ for all $s\in Act^*$, we have $Y_n\subseteq Y$ from the definition of $Y$. So $Y_n=Y$. $\Box$

Next we prove the decidability of $\A\models\U f$. Indeed, we can prove the following lemma from which it follows that $Y(\A,\U,f)=Y(\A,\G,f)$.
\begin{lem}
$\A\models\U f$ iff $\hs_{ini}\subseteq Y(\A,\G,f)$.
\end{lem}
\textit{Proof:} The ``if'' part can be verified by observation of $$\A\models\G f \Rightarrow \A\models\U f.$$ So we only need to prove the ``only if'' part. We assume $\A\models\U f$. Then for any $\ket{\psi_0}\in\hs_{ini}$, we have $\A(\psi_0)\models\U f$. According to clause 4) of Lemma~\ref{lem:relation}, we know that $Act^*-\Lan(\A(\psi_0),f)$ is finite. Then there exists some integer $N\geq 0$ such that $s\in \Lan(\A(\psi_0),f)$ whenever $|s|\geq N$. We choose $s=\alpha^N$. Then $U_\alpha^N\ket{\psi_0}\in Y$ for any $\alpha\in Act$. Note that $U_\alpha Y=Y$, we have $\ket{\psi_0}\in U_\alpha^{-N}Y=Y$. $\Box$

\section{Conclusion}\label{sec:con}
We have investigated the decision problem of quantum reachability: decide whether or not a set of quantum states is reachable by a quantum system modelled by a quantum automaton. The reachable sets considered in this paper are defined as boolean combinations of (or described by classical propositional logical formula over) the set of (closed) subspaces of the state Hilbert space of the system. Four types of reachability properties have been studied: eventually reachable, globally reachable, ultimately forever reachable, and infinitely often reachable.
Our major contribution is the (un)decidable results:
\begin{itemize}
\item All of these four reachability properties are undecidable even for a certain class of the reachable sets which are formalized by logical formulas of a simple form;
\item Whenever the reachable set is a union of finitely many subspaces, the problem is decidable for globally reachable, ultimately forever reachable and infinitely often reachable. In particular, it is decidable when the reachable set contains only finitely many quantum states.
\end{itemize}
One of our main proof techniques is to demonstrate that quantum reachability problem is a generalization of the Skolem's problem for unitary matrices. The undecidable results for global reachability, ultimately forever reachability and infinitely often reachability have been derived directly by employing the undecidability of a relevant emptiness problem. Nevertheless, the celebrated Skolem-Mahler-Lech theorem has been applied to the development of algorithms showing the decidable results. Another technique we have employed is to encode a 2-counter Minsky machine using a quantum automaton. It was used to prove undecidability of the eventually reachable property. This approach is interesting, since it provides a new way to demonstrate quantum undecidability other than reduction from the PCP that has been the main technique for the same purpose in previous works.

The problem whether or not $\A\models\F f$ is decidable for $\|f\|$ being a finite union of subspaces has been left unsolved. In fact, this problem is difficult even for a very special case where $|Act|=1$ and $\|f\|$ is a single subspace. We have shown that such a reachability problem is equivalent to the emptiness Skolem's problem~\ref{prob:empty} for $\{a_n\}_{n=0}^\infty$ defined by Eq.~(\ref{equ:matrix}) with $M$ being a unitary operator. Unfortunately, the emptiness Skolem's problem is still open even for $n=5$~\cite{QW12}.

The model of quantum systems used in this paper is quantum automata. Another problem for further studies is (un)decidability of the reachability properties considered in this paper for a more general model, namely quantum Markov chains \cite{YFYY13} where actions can be not only unitary transformations but also super-operators.

\section*{Appendix}
\subsection*{A. Proof of Lemma~\ref{lem:period-p}} We algorithmically construct a positive integer $p$ satisfying the following condition: for any two eigenvalues $\lambda$ and $\mu$ of $U_\alpha$, if $(\lambda/\mu)^n=1$ for some integer $n$, then $(\lambda/\mu)^p=1$. Note that all roots of the characteristic polynomial $f(x)$ of $U\otimes U^\dagger$ are exactly all quotients $\lambda/\mu$ of two eigenvalues of $U$. If for some quotient and integer $n$, $(\lambda/\mu)^n=1$, we let $n$ be the minimal positive integer number satisfying this condition. Then $\lambda/\mu$ should also be a root of the $n$th cyclotomic polynomial $\Phi_n(x)$. Thus $\Phi_n(x)$ should be a divisor of $f(x)$ since $\Phi_n(x)$ is irreducible. Therefore, all of such $n$'s can be obtained by checking whether or not $\Phi_n(x)|f(x)$. Finally, we put $p$ to be the least common multiple of them. It is easy to verify that $(\lambda/\mu)^p=1$ for all such quotients.

Now we prove that $p$ is really what we want. Suppose $U^nK=K$, then there exists a basis of $K$ such that all states of this basis are eigenstates of $U^n$. It suffices to prove that any eigenstate of $U^n$ is
also an eigenstate of $U^p$. Now we prove it by showing that any eigenspace $W$ of $U^n$ is also an eigenspace of $U^p$. Since all eigenstates of $U$ are eigenstates of $U^n$, we can choose a set of eigenstates of $U$ to form a basis of $W$. Consider any two of these states, written as $\ket{\psi}$ and $\ket{\phi}$, and written as $\lambda$ and $\mu$, respectively, for the corresponding eigenvalues of $U$. Then we have $(\lambda/\mu)^{n}=1$, and according to our choice of $p$, $(\lambda/\mu)^p=1$. So $\ket{\psi}$ and $\ket{\phi}$ are in the same eigenspace of $U^p$. As these two states are arbitrarily chosen, it implies that all of states in this basis of $W$ are in the same eigenspace of $U^p$. Thus $W$ is an eigenspace of $U^p$. $\Box$

\subsection*{B. Correctness of Algorithm \ref{alg:check1}} For any $q\in\N$, we write $K_q$ as the maximal subspace of $V$ such that $U_\alpha^qK_q=K_q$. Then $K_p=K=K(U_\alpha,V)$. We prove that $K_q$ can be characterized as the following set of sates:
$$\{\ket{\psi}\in V|\forall n\in\N,\ U_\alpha^{qn}\ket{\psi}\in V\}.$$
In fact, it is easy to verify that any state in $K_q$ is also in this set. On the other hand, for any state $\ket{\psi}$ in this set, $\spa\{U_\alpha^{qn}\ket{\psi}|n=0,1,\cdots\}$ is both a subspace of $V$ and an invariant subspace of $U_\alpha^q$, so it is a subspace of $K_q$ according to the maximality of $K_q$. Then $\ket{\psi}\in K_q$. Therefore $K_q$ is equal to the set.

Now for each $|\psi\rangle\in\hs$, according to clause 2) in Lemma~\ref{lem:relation}, $\A(\psi)\models\I V$ iff $\Lan(\A(\psi),V)=\{n\geq 0|U_\alpha^n\ket{\psi}\in V\}$ satisfies liveness condition Eq.~(\ref{equ:liveness-condi}), namely it is infinite in this case. Note that $U_\alpha^n\ket{\psi}\in V$ iff $\tr(P_{V^\perp}\Uo^n(\psi))=0$, where $\psi$ is the density operator of $\ket{\psi}$, $P_{V^\perp}$ is the projection operator of $V^\perp$ and $\Uo$ is the super-operator of $U_\alpha$. Since $\{\tr(P_{V^\perp}\Uo^n(\psi))\}_{n=0}^\infty$ is a linear recurrence sequence, according to Theorem~\ref{the:SML}, $\Lan(\A(\psi),V)$ is semi-linear, and thus it is infinite if and only if it contains an arithmetic progression $\{qn+r\}_{n=0}^\infty$. Then
\begin{equation}\label{uuu}\begin{split}Y &=\{|\psi\rangle|\Lan(\A(\psi),V)\ {\rm is\ infinite}\}\\ &=\{\ket{\psi}|\exists q,r\in\N.\forall n\in\N.\ U_\alpha^{qn+r}\ket{\psi}\in V\}\\
&=\{|\psi\rangle|\exists q,r\in\N.U_\alpha^r|\psi\rangle\in K_q\}\\ &=\{|\psi\rangle|\exists q,r\in\N.|\psi\rangle\in U_\alpha^{q-r}K_q\}\\
&=\bigcup_{q,r\geq 0}U_\alpha^rK_q=\bigcup_{r=0}^\infty U_\alpha^rK_p=\bigcup_{r=0}^{p-1}U_\alpha^rK.
\end{split}\end{equation}
The last two equalities in Eq.~(\ref{uuu}) come from the following observation. For each integer $q$, since $U_\alpha^{q}K_q=K_q$, by Lemma~\ref{lem:period-p} we have $U_\alpha^pK_q=K_q$. Thus $K_q\subseteq K_p=K$ follows from maximality of $K$. $\Box$

\subsection*{C. Proof of Lemma~\ref{lem:proper-W}}
We need to consider the two following cases:
\begin{itemize}
\item Case 1. Condition 1) in Lemma \ref{lem:3-condi} is not satisfied by $X$.
\item Case 2. Condition 1) in Lemma \ref{lem:3-condi} is satisfied by $X$ but condition 2) is not.\end{itemize}

\textit{Proof for case 1:} Since condition 1) is not satisfied, we can find all $Y_i$ and $\alpha\in Act$ such that $U_\alpha Y_i$ is not any $Y_j$. We choose $Y_i$ with the maximal dimension and claim that for any $\alpha\in Act$, $U_\alpha Y_i$ can not be included in any $Y_j$. Otherwise, $U_\alpha Y_i$ is a proper subspace of some $Y_j$, and $\dim Y_j>\dim Y_i$. It is easy to prove by induction on $n$ that all the subspaces $U_\alpha^n Y_j\ (n=0,1,\cdots)$ are in $\{Y_1,Y_2,\cdots,Y_m\}$. So, there exists some $n_1$ and $n_2$ such that $n_2>n_1$ and $U_\alpha^{n_1} Y_j=U_\alpha^{n_2} Y_j$. Then $Y_i$ is a proper subset of $U_\alpha^{-1} Y_j=U_\alpha^{n_2-n_1-1} Y_j$, which is in $\{Y_1,Y_2,\cdots,Y_m\}$. This contradicts to the assumption that any two subspaces in $\{Y_1,Y_2,\cdots,Y_m\}$ do not include each other.

Now we choose $W_j=Y_i\cap U_\alpha^{-1} Y_j\ (j=1,2,\cdots m)$ for $Y_i$. All of these are proper subspaces of $Y_i$. On the other hand, from the definition of $Y$, one can easily verify that
$U_\alpha\ket{\psi}\in Y$ for all $\ket{\psi}\in Y$ and for all $\alpha\in Act.$
Then for any state $\ket{\psi}\in Y\cap Y_i$, we know that $U_\alpha\ket{\psi}\in Y\subseteq\cup X$. So $\ket{\psi}$ is in some $U_\alpha^{-1}Y_j$, and thus $\ket{\psi}\in Y_i\cap U_\alpha^{-1} Y_j=W_j$. Then Eq.~(\ref{equ:included}) holds. $\Box$

To prove Lemma~\ref{lem:proper-W} for case 2, we need the following:
\begin{lem}\label{lem:small}
For any $\ket{\psi_0}\in Y$ and $\alpha_1,\alpha_2,\cdots,\alpha_k\in Act$, there exists some $r\in\{0,\cdots,k-1\}$, some $t\in\{1,2,\cdots,m\}$, and some $n\in\N$, such that
\begin{equation}\label{equ:circle}
\ket{\psi_0}\in U_{\alpha_1}^{-1}U_{\alpha_2}^{-1}\cdots U_{\alpha_r}^{-1}T^n K(T,V_t),
\end{equation}
where $T=U_{\alpha_{r+1}}\cdots U_{\alpha_k}U_{\alpha_1}\cdots U_{\alpha_r}$, and $K(T,V_t)$ is defined as in Algorithm~\ref{alg:check1}.
\end{lem}
\textit{Proof:} We consider the path $p$ of repeatedly performing $U_{\alpha_1},U_{\alpha_2},\cdots, U_{\alpha_k}$ from the initial state $\ket{\psi_0}$:
\begin{equation}\begin{split}
p=&\ket{\psi_0}\overset{U_{\alpha_1}}\rightarrow\ket{\psi_1}\overset{U_{\alpha_2}}\rightarrow\cdots\overset{U_{\alpha_{k-1}}}\rightarrow\ket{\psi_{k-1}}\overset{U_{\alpha_k}}\rightarrow\\
&\ket{\psi_k}\overset{U_{\alpha_1}}\rightarrow\ket{\psi_{k+1}}\overset{U_{\alpha_2}}\rightarrow\cdots\overset{U_{\alpha_{k-1}}}\rightarrow\ket{\psi_{2k-1}}\overset{U_{\alpha_k}}\rightarrow\\
&\cdots.
\end{split}\end{equation}
Then $\ket{\psi_{kn+r+1}}=U_{\alpha_{r+1}}\ket{\psi_{kn+r}}$, for all $n\in\N$ and $r\in\{0,\cdots,k-1\}$.
Since $\sigma(p)\models\I f$, we have $\ket{\psi_n}\in \|f\|$ for infinitely many $n$. This further implies that there exists some $r\in\{0,1,\cdots,k-1\}$ and some $t\in\{1,2,\cdots,m\}$ such that $\ket{\psi_{kn+r}}\in V_t$ for infinitely many $n$.
We put $T=U_{\alpha_{r+1}}\cdots U_{\alpha_k}U_{\alpha_1}\cdots U_{\alpha_r}$. Then the set $\{n|T^n\ket{\psi_r}\in V_t\}$ is infinite. According to the result of single unitary case, we have $\ket{\psi_r}\in T^n K(T,V_t)$. This is exactly Eq.~(\ref{equ:circle}). $\Box$

Now we are able to prove Lemma~\ref{lem:proper-W} for case 2.

\textit{Proof for case 2:} Since the condition 1) is satisfied but condition 2) is not, we can find a simple loop $$Y_{r_0}\overset{U_{\alpha_1}}\rightarrow Y_{r_1}\overset{U_{\alpha_2}}\rightarrow \cdots\overset{U_{\alpha_{k-1}}}\rightarrow Y_{r_{k-1}}\overset{U_{\alpha_k}}\rightarrow Y_{r_0},$$ such that $Y_{r_i}\nsubseteq V_t$ for all $i\in\{0,1,\cdots,k-1\}$ and all $t\in\{1,2,\cdots,m\}$. We choose $Y_{r_0}$ and construct $W_1,W_2,\cdots,W_l$ to be proper subspaces of it. In fact, for each $i$, we write $T_i=U_{\alpha_{i+1}}\cdots U_{\alpha_k}U_{\alpha_1}\cdots U_{\alpha_i}$. It holds that $T_i^nY_{r_i}=Y_{r_i}\nsubseteq V_t$, and $Y_{r_i}\nsubseteq T_i^nK(T_i,V_t)$ for all integer $n$. Put $$R_{i,t,n}=U_{\alpha_1}^{-1}U_{\alpha_2}^{-1}\cdots U_{\alpha_i}^{-1}T_i^nK(T_i,V_t),$$ then it actually means $Y_{r_0}\nsubseteq R_{i,t,n}$. Note that $T_i^{p_i}K(T_i,V_t)=K(T_i,V_t)$ for the period $p_i$ of $T_i$. So the set $\{R_{i,t,n}|n=0,\pm 1,\pm 2,\cdots\}$ is a finite set for any $i=0,1,\cdots,k-1$ and any $t=1,2,\cdots,m$. Therefore, we can choose $W_1,W_2,\cdots,W_l$ to be all of the $Y_{r_0}\cap R_{i,t,n}$'s.
The condition of Eq.~(\ref{equ:included}) can be easily verified, since for any state $\ket{\psi}\in Y\cap Y_{r_0}$, we have $\ket{\psi}$ is in some $R_{i,t,n}$ according to Lemma~\ref{lem:small}. $\Box$


\begin{thebibliography}{99}

\bibitem{AT12} C. Altafini, and F. Ticozzi, Modeling and control of quantum systems: An introduction, \textit{IEEE Transactions on Automatic Control}, 57(2012)1898.

\bibitem{AI99} M. Amano, and K. Iwama. Undecidability on quantum finite automata, in: \textit{Proceedings of the thirty-first annual ACM symposium on Theory of computing (STOC)}, 1999 pp. 368-375.

\bibitem{AGN13} E. Ardeshir-Larijani, S. J. Gay and R. Nagarajan, Equivalence checking of quantum protocols, in: \textit{Proceedings of the 19th International Conference on Tools and Algorithms for the Construction and Analysis of Systems (TACAS)}, Springer LNCS 7795, 2013, pp. 478-492.

\bibitem{BK08} C. Baier and J. -P. Katoen, \textit{Principles of Model Checking}, MIT Press, Cambridge, Massachusetts, 2008.

\bibitem{BM76} J. Berstel and M. Mignotte, Deux propri\'{e}t\'{e}s d\'{e}cidables des suites r\'{e}currentes
lin\'{e}aires, \textit{Bull. Soc. Math. France}, 104(1976)175-184.

\bibitem{BN36} G. Birkhoff and J. von Neumann, The Logic of Quantum Mechanics, \textit{Annals of Mathematics}, 37(1936)823-843.

\bibitem{BJKP05} V. D. Blondel, E. Jeandel, P. Koiran and N. Portier, Decidable and undecidable problems about quantum automata, \textit{SIAM Journal on Computing}, 34(2005)1464-1473.

\bibitem{CK98} J. Cassaigne and J. Karhum\"{a}ki: Examples of undecidable problems for
2-generator matrix semigroups, \textit{Theoretical Computer Science} 204(1998)29-34.

\bibitem{CZ12} J. I. Cirac and P. Zoller, Goals and opportunities in quantum simulation, \textit{Nature Physics}, 8(2012)264-266.

\bibitem{EMG12} J. Eisert, M. P. M\"{u}ller and C. Gogolin, Quantum measurement occurrence is undecidable, \textit{Physcal Review Letters}, 108(2012)260501.

\bibitem{GPN10} S. J. Gay, R. Nagarajan, and N. Papanikolaou, Specification and verification of quantum protocols, in: \textit{Semantic Techniques in Quantum Computation} (S. J. Gay and I. Mackie, eds.), Cambridge
University Press, 2010, pp. 414-472.

\bibitem{GV13} A. S. Green, P. L. Lumsdaine, N. J. Ross, P. Selinger and B. Valiron, Quipper: A scalable quantum programming language, in: \textit{Proceedings of the 34th ACM Conference on Programming Language Design and Implementation (PLDI)}, 2013, pp. 333-342.

\bibitem{HHHK05} V. Halava, T. Harju, M. Hirvensalo, and J. Karhum\"{a}ki, \textit{Skolem's Problem: On
the Border between Decidability and Undecidability}, Technical Report 683, Turku
Centre for Computer Science, 2005.

\bibitem{KW97} A. Kondacs and J. Watrous, On the power of quantum finite state automata, in: \textit{Proc. 38th Symposium on Foundation of Computer Science (FOCS)}, 1997, pp. 66-75.

\bibitem{Lech52} C. Lech, A note on recurring series, \textit{Ark. Mat.} 2(1953)417-421.

\bibitem{LYY12} Y. J. Li, N. K. Yu and M. S. Ying, Termination of nondeterministic quantum programs, \textit{Acta Informatica}  (published online October 2013; also short presentation of LICS'2012).

\bibitem{Mahler35} K. Mahler, Eine arithmetische eigenschaft der Taylor koeffizienten rationaler
funktionen, in: \textit{Proc. Akad. Wet. Amsterdam}, 38, 1935.

\bibitem{Min76} M. L. Minsky, \textit{Computation: finite and infinite machines}, Prentice-Hall, 1967.

\bibitem{NC00} M. A. Nielsen and I. L. Chuang, \textit{Quantum Computation and Quantum Information}, Cambridge University Press, 2000.

\bibitem{Paz71} A. Paz, \textit{Introduction to probabilistic automata}, Academic Press, New York, 1971.

\bibitem{Pos46} E. L. Post, A variant of a recursively unsolvable problem, \textit{Bulletin of the American
Mathematical Society}, 52(1946)264-268.

\bibitem{QW12} J. Ouaknine and J. Worrell, Decision Problems for Linear Recurrence Sequences, in: \textit{Reachability Problems}, Springer LNCS 7550, 2012, pp. 21-28.

\bibitem{SS78} A. Salomaa and M. Soittola, \textit{Automata-Theoretic Aspects of Formal Power
Series}, Springer-Verlag, 1978.

\bibitem{Skolem34} T. Skolem, Ein verfahren zur behandlung gewisser exponentialer gleichungen, in: \textit{ Proceedings of the 8th Congress of Scandinavian Mathematicians}, Stockholm, 1934, pp. 163-188.

\bibitem{SSL02} S. G. Schirmer, A. I. Solomon and J. V. Leahy, Criteria for reachability of quantum states, \textit{Journal of Physics A: Mathematical and General}, 35(2002)8551-8562.

\bibitem{Yin11} M. S. Ying, Floyd-Hoare logic for quantum programs, \textit{ACM Transactions on Programming Languages and Systems}, (2011) art. no. 19.

\bibitem{YYFD13} M. S. Ying, N. K. Yu, Y. Feng, and R. Y. Duan, Verification of quantum programs, \textit{Science of Computer Programming}, 78(2013)1679-1700.

\bibitem{YFYY13} S. G. Ying, Y. Feng, N. K. Yu and M. S. Ying, Reachability probabilities of quantum Markov chains, in: \textit{Proceedings of the 24th International Conference on Concurrency Theory (CONCUR)}, Springer LNCS 8052, 2013, pp. 334-348.

\bibitem{YY12} N. K. Yu and M. S. Ying, Reachability and termination analysis of concurrent quantum programs, in: \textit{Proceedings of the 23rd International Conference on Concurrency Theory (CONCUR)}, Springer LNCS 7454, 2012, pp. 69-83.
\end{thebibliography}
\end{document}